\documentstyle[preprint,aps,prb,psfigure]{revtex}

\begin{document}
 
\newcommand{\beq}{\begin{equation}}
\newcommand{\eeq}{\end{equation}}
\draft

\title{The magnetic form factor of the neutron}
\author{E.E.W. Bruins$^{a,b}$}
\address{ $^a$ Laboratory for Nuclear Science, 
Massachusetts Institute of Technology, \\
77 Massachusetts Avenue, Cambridge, MA 02139, USA}
\address{$^b$ Temporary Address: DESY, Notkestrasse 85, 22607 Hamburg, Germany}

\maketitle

\begin{abstract}  
A review of neutron form factor measurements is given. We focus on recent
measurements of the neutron magnetic form factor $G_M^n$, and discuss in 
detail our measurements of this quantity at momentum transfers $Q^2= 0.1 - 
0.6$~(GeV/c)$^2$.
\end{abstract}
 


\narrowtext


\section{Nucleon form factors}

In the one photon exchange approximation, the elastic electron -- nucleon
differential cross section is described by the so-called Rosenbluth formula
\cite{rosenbluth}. 
The expression contains two electromagnetic form factors which depend on 
the four momentum transfer squared $q_{\mu}^2 = -Q^2$ in the reaction, only.
These form factors, usually called $G_E(Q^2)$ (electric) and $G_M(Q^2)$ 
(magnetic), can be thought of as the Fourier representation of the electric 
charge and current distributions in the nucleon. 
This relation, however, does not hold at realistic $Q^2$, where the recoil 
of the nucleon and relativistic kinematics are non-negligible. 
In the Breit (or brick wall) frame, which is the kinematical frame in which 
no energy is transferred, $G_E$ and $G_M$ are proportional to the nucleon 
charge and current matrix elements, respectively.

In the late 1950s, Hofstadter and collaborators initiated a series of 
experiments in which the form factors of the proton and the neutron
were measured~\cite{devries}. 
The proton form factors were obtained from the measurement of the elastic 
electron--proton ($e-p$) cross sections at two different kinematics, 
keeping $Q^2$ fixed.
Since no free neutron targets are available, the neutron form factors were
obtained by measuring the quasi-elastic electron--deuteron ($e-d$) cross 
sections at the same kinematics, and subsequently subtracting the contribution 
from the proton using the $e-p$ data.

Most measurements of the proton and neutron form factors, even recent 
ones~\cite{bosted,walker,lung}, have used this very same technique. 
By now, the proton form factors are conventionally considered to be known 
accurately: The proton magnetic form factor $G_M^p$ is known with an accuracy 
of 2 \% up to at least $Q^2=1$~(GeV/c)$^2$. 
The electric form factor $G_E^p$ is known to about 5 \% at best, due to the 
fact that 
its numerical value is smaller than $G_M^p$ in the accessible range of momentum 
transfers. 
In the most elaborate recent work on neutron form factors from inclusive 
measurements, Lung {\it et al.}~\cite{lung} have shown that the neutron 
magnetic form factor $G_M^n$ can be extracted with an accuracy of 10 -- 15 \% 
using this method. 
The neutron electric form factor $G_E^n$ is essentially undetermined, and 
recent studies~\cite{spin96-gen-reference} suggest that polarization variables 
might be more useful for an unambiguous determination of this small quantity.

Several measurements have attempted an accurate determination of $G_M^n$ by 
employing coincident neutron detection, initiated in 1962 by the measurement
of Stein {\it et al.}~\cite{stein62}.
Neutron detection involves a dedicated neutron detector, usually built from
scintillating material, and a careful calibration of its detection efficiency. 
Because of these experimentally non-trivial considerations, the early 
measurements~\cite{stein62,stein63,bartel69,bartel72} did not yield both
reliable and accurate data, and consequently large systematic uncertainties 
had to be introduced.
Specific problems were the position dependence of the neutron detection 
efficiency, which was essentially unknown~\cite{stein63}, and the large proton 
contribution which forced the experimenters to use heavy 
shielding~\cite{bartel69,markowitz} or to measure only at high momentum
transfers~\cite{bartel72}.

With the advent of high duty cycle electron machines (ELSA (Bonn), 
MAMI (Mainz), Jefferson Laboratory (Newport News)) accurate determinations 
of the neutron form factors are considered realistic, which has lead to a 
renewed interest in these quantities.

\section{Determination of the neutron magnetic form factor}

We have determined the magnetic form factor of the neutron, $G_M^n$, 
at four different momentum transfers between $Q^2 = 0.1$ and 0.6 (GeV/c)$^2$. 
The goal of the measurement was to minimize the experimental and theoretical 
uncertainties and to deliver the first accurate determination of the 
behavior of $G_M^n$ over an appreciable range of $Q^2$.

The ratio of neutron and proton yields at quasifree kinematics was 
measured for the reactions $^2$H(e,e'n) and $^2$H(e,e'p) at 
momentum transfers $Q^2$=0.125, 0.255, 0.417 and 0.605~(GeV/c)$^2$
(Kin. I, II, III and IV), 
detecting the neutron and the proton simultaneously in the same 
scintillator array, the nucleon detector. The neutron detection efficiency 
was measured {\it in situ} with the $^1$H($\gamma$,$\pi^+$)n reaction.
From this, the ratio $R$ of $^2$H(e,e'n) and 
$^2$H(e,e'p) cross sections was determined and used to extract the 
neutron magnetic form factor $G_M^n$ in a model insensitive approach, 
resulting in an inaccuracy between 2.1 and 3.3~\% in $G_M^n$.

The method employed, first suggested by Durand~\cite{durand}, was 
previously applied in only a few experiments~\cite{bartel69,bartel72}. 
In this way 
the luminosity, the electron detection efficiency, and the electron solid 
angle cancel out, while the nucleon acceptances and the choice of the 
deuteron wave function cancel out to first order. The 
lay-out of the experiment minimized corrections due to proton losses,
and the restriction to quasifree kinematics minimized nuclear effects 
and their corresponding uncertainties. 

The ratio $R$ of the cross sections of the $^2$H(e,e'n) and $^2$H(e,e'p) 
reactions was determined after taking into account proton losses due to 
nuclear reactions, multiple scattering and edge effects, and an 
{\it in situ} calibration of the neutron detection efficiency. 
$G_M^n$ was extracted taking into account nuclear effects, using the 
known $e-p$ cross sections and available information on $G_E^n$. 
$G_E^n$ contributes less than 2 \% to the $e-n$ cross section and is 
not a dominant source of uncertainty.

\subsection{Experimental details}

\subsubsection{The detector setup}

The electron beam (between 900 and 1600 MeV, 20 to 60 nA, 20 to 50 \%
duty cycle, depending on kinematics), delivered by the ELSA accelerator at 
the Physics Institute at Bonn, impinged axially on a cylindrical 
target (length 10~cm) made from 125~$\mu$m thick kapton and
filled with liquid Hydrogen or Deuterium. 
Electrons (and pions from the $^1$H($\gamma$,$\pi^+$)n reaction)
were detected in the ELAN magnetic spectrometer with four 
scintillators and MWPCs and one \v Cerenkov counter.

Protons and neutrons were detected in the nucleon detector using the time 
of flight method.  
Since protons are about hundred times more abundant than neutrons, 
a reliable particle identification is important. 
We used three thin veto scintillators which allowed 
a misidentification rate of far less than $10^{-4}$, 
and the verification of the losses due to pile-up \cite{joosse}.

For Kinematics I, II and III, the nucleon detector, which was used previously,
consisted of a total of five scintillators (NE102a) of
dimensions $25~\times~25~$cm$^2$ with a thickness of 2~mm for the $\Delta E$ 
detectors (1~mm for Kin. I) and 50~mm for the $E$ counters.  
The first $E$ counter ($E_{front}$) detected simultaneously 
protons and neutrons, while shielding behind $E_{front}$ prevented
protons from the quasielastic $^2$H(e,e'p) reaction to reach $E_{back}$. The 
neutrons detected in $E_{back}$ were used to determine the losses in the 
neutron yield in $E_{front}$ caused by the software gates on the veto 
detectors.  For Kin.~IV, the detectors of dimensions 100~$\times$~18~cm$^2$
and thickness 1~cm and 18~cm, respectively,
had a two-sided readout. 
At this large proton energy the use of lead to shield $E_{back}$ 
becomes impracticable. We therefore installed two
veto detectors between $E_{front}$ and $E_{back}$ in order to 
determine the neutron losses. For a more detailed description, see 
Ref.~\cite{bruins}.
Each measurement of $R$ was bracketed by two calibration runs for the neutron 
detection efficiency.  To switch from the calibration to the measurement of 
the ratio, only the target and the spectrometer settings needed to be changed, 
which reduced the risk of systematic errors.

\subsection{The analysis}
\label{thissec}
Neutrons detected in the nucleon detector were defined by a signal in 
$E_{front}$ or in $E_{back}$ at the correct time with respect to the 
spectrometer and in excess of a certain software adjustable threshold, 
together with a neutron condition on the veto counters.  Different veto 
conditions on the ADC and TDC information were applied and used for stringent 
cross checks.  Thanks to the duty cycle of the ELSA stretcher ring (20
to 50 \%), a signal to noise ratio of 200:1 for neutrons was achieved 
(Fig.~\ref{figure:tof}).

The stability of the neutron detection efficiency of $E_{front}$, $\eta_n$,
was monitored by means of the ADC signal for protons, and by 
the number of neutrons coming from the $^2$H(e,e'n) reaction scaled 
with the number of electrons in the spectrometer.  At Kin. III and IV,
the $^2$H($\gamma$,p)n reaction served as an additional monitor
for the stability of $\eta_n$.
This reaction can not be used to determine the absolute value of $\eta_n$
since pion production cannot be excluded.

The reaction $^1$H($\gamma$,$\pi^+$)n was used to obtain an $in$ $situ$, 
absolute determination of the detector efficiency for neutrons. 
Under the assumption that only real photons are used, the kinematics of 
this reaction can be arranged in such a way as to unambiguously 
tag neutrons impinging on the nucleon detector centered at the same energy as 
those from the $^2$H(e,e'n) reaction.
Eighty percent of the pion flux in the spectrometer was determined to 
actually originate from virtual photons, i.e. from the $^1$H(e,$\pi^+$)e'X 
reaction. 
Monte Carlo simulations~\cite{reike}, however, show that more than 99 \% of 
the pions which enter the spectrometer 
(and satisfy the cuts used in the analysis, as described in Ref.~\cite{bruins}) 
originate from electrons scattering to extremely small angles ($\ll 5^o$). 
Because of the stringent requirements put on the pion kinematics, 
to resulting data set can be treated as originating from the
real photon induced $^1$H($\gamma$,$\pi^+$)n reaction.
This consideration is unfortunately not supported by simulations
performed by the Basel group, which have recently published competing
$G_M^n$ results~\cite{thissymp}. The discrepancy has to be resolved in
the near future. Curiously, 
for neutrons with an average energy of 61~MeV, the measured
detection efficiency in the center of the nucleon detector as a function of 
threshold agreed with the one established earlier \cite{joosse,anklin}.

In order to obtain $R$, the measured yields must be corrected for: $i)$~the 
net proton losses due to nuclear reactions and multiple scattering 
in the material between the reaction vertex and the detector;
$ii)$~the dependence of the neutron detection efficiency on the distribution 
of the neutrons in space and energy over the detector surface. Details about
the small corrections for Hydrogen contamination of the Deuterium, and for the 
contributions of the target end caps to the different reactions can be found 
in Ref.~\cite{bruins}.  In addition, for an extraction of $G_M^n$ one must 
evaluate nuclear effects, such as final state interactions (FSI), meson 
exchange currents (MEC) and isobar currents (IC), which alter the 
proton and neutron yields expected from free proton respectively neutron
targets. 

The $^1$H(e,e')p reaction was used for the determination of the 
proton detection efficiency, including the above mentioned losses.
The experimentally found losses were consistent with 
numerical checks done with the GEANT package \cite{geant} which was 
extended to include total proton cross sections at low energies \cite{zisis}.
Protons from 
the $^1$H(e,e')p and the $^2$H(e,e'p)n reaction also were used to calibrate 
and monitor the light response of the scintillators in view of the threshold 
dependence of the neutron detection efficiency. 

The reactions $^2$H(e,e'n) and $^1$H($\gamma$,$\pi^{+}$)n lead to different 
energy and position distributions of neutrons in $E_{front}$. While the 
neutron distribution for the $^1$H($\gamma$,$\pi^{+}$)n reaction is known from 
experiment, the $^2$H(e,e'n) distribution was obtained using the ENIGMA Monte
Carlo code \cite{janv}.  The second order effect of the different 
distributions (including edge effects) on the ratio $R$ was simulated with the 
program KSUVAX \cite{cecil}.
The response of the detector used for Kin. I, II 
and III had been calibrated carefully previously \cite{loppacher} using 
a tagged neutron beam at PSI. The positional dependence of the response of the 
detector used for Kin. IV was studied by means of its double read out system.

\subsection{Nuclear effects}

Nuclear effects cause $R$ to differ from the 
expected ratio $R_{free}$ for free nucleons: $R_{free} = R \ast 
(1 - \delta R)$.  The corrections $\delta R$ were calculated in two different 
models, of which the first \cite{arensugg} includes FSI, MEC and IC, and 
shows that FSI dominate the other corrections, whereas the second one 
\cite{tjonsugg} includes only FSI but is fully relativistic. 
It appears that the result for $G_M^n$ obtained from $R$ is affected to 
less than 0.4 \% by relativistic effects. 
We use the
difference between the two calculations, including FSI only, as an
indication of the inherent model uncertainties, and their average as
the most probable correct value. 
To this average, we add then the effects of MEC and IC which are taken 
as the difference between the full calculation, and the calculation
including FSI only, both within the same model~\cite{arensugg}.
We conservatively estimate the uncertainty due to these effects to be half their size.
The values of $\delta R$ range from 2.2 to 8.5 \%, depending on the kinematics.

\subsection{Results}

For the extraction of $G_M^n$, the $e-p$ cross sections are needed. 
The measured cross section data~\cite{jourdan} have been averaged 
in small intervals around the central $Q^2$ values,
taking into account the $Q^2$ dependence obtained from the
dipole form factors for the proton ($G_E^p$=$G_D$=(1+$Q^2$/$a^2$)$^{-2}$, 
$G_M^p$=2.793$\ast G_E^p$, and $a^2$=0.710 (GeV/c)$^2$). 
The statistical errors of the used data points determine 
the statistical error in the average. In order 
to obtain the systematical error, the average has been recalculated after each 
$e-p$ data set had been changed individually by its systematical errors, and 
adding the resulting changes quadratically. 
$G_E^n$ was chosen to be 0.037~$\pm$~0.017~\cite{platchkov}. 

\begin{figure}
\strut\psfigure[4.0cm]{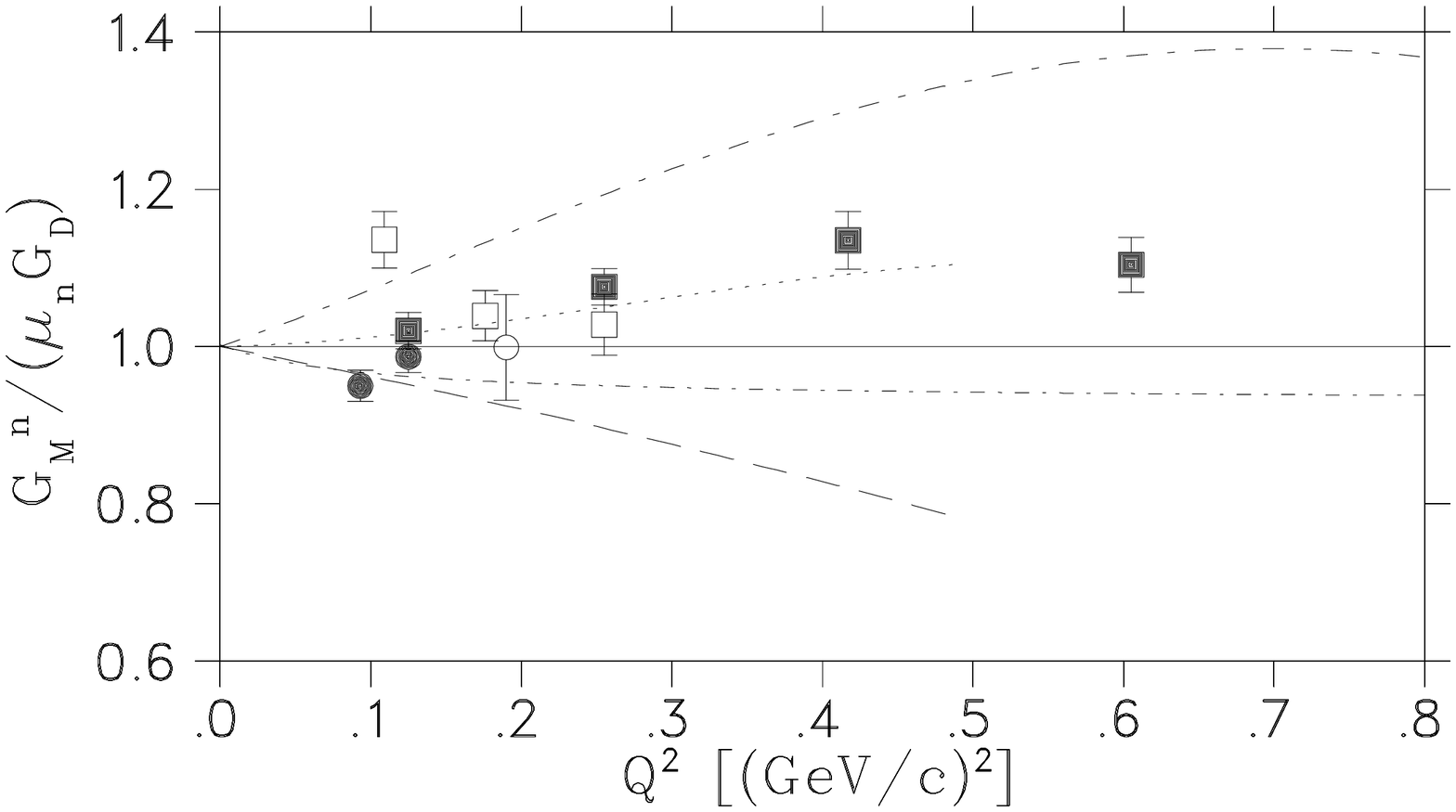}
\caption{
Data from recent $G_M^n$ measurements up to $Q^2$=0.8 (GeV/c)$^2$,
scaled to the dipole fit:
black circles [17], open squares [11], open circle [25].
The black squares are from this work. Several model predictions are
shown. VMD: The minimal (dotted line) and complete 
(long dash) model of [28]. Constituent Quark Model: 
non-relativistic calculation of [29] (dot-long dash),
relativistic calculation of [30] (dot-short dash).}
\label{figure:data}
\end{figure}

The results for $G_M^n$ are shown in Fig.~\ref{figure:data}.
We confirm our pilot measurement at $Q^2$=0.125~(GeV/c)$^2$, performed
at the MEA accelerator (NIKHEF, Amsterdam) with the same nucleon detector
as used in Kin. I, II and III. Note that the {\it published} results differ by
more than one standard deviation, only because of the difference in the value
for the proton cross section used. This difference, 2 \%, comes from the 
different $Q^2$ range of the $e-p$ data used in the fit and reflects our 
knowledge of the $e-p$ cross section.
Our value for $G_M^n$ at this momentum transfer is significantly higher than  
our previous result for $Q^2$=0.093 (GeV/c)$^2$ obtained at MEA.
Fig.~\ref{figure:data} suggests a slope in $G_M^n$/$G_D$ around these momentum
transfers.

Recently Markowitz et al. 
\cite{markowitz} determined $G_M^n$ from an absolute measurement of the 
$^2$H(e,e'n) cross section. At 0.109~(GeV/c)$^2$, their result lies 13.6~\% 
above the dipole fit, whereas our result at 0.125 (GeV/c)$^2$
is only 2.5~\% above.  Relativistic effects, which have not been applied in 
Ref.\cite{markowitz}, do not appear to remove this discrepancy.  
Our results agree with the other results of Markowitz, 
and with the result obtained by Gao \cite{gao} using polarisation techniques.
Fig.~\ref{figure:data} shows that most of the recent theoretical
predictions of $G_M^n$ are not in agreement with our data. We only show
some of the (rare) theoretical predictions which are not fits to
$G_M^n$ data. The only model which reproduces the trend indicated by the
data is the minimal model of Meissner~\cite{meissner}. Since Meissner's
{\it complete} model does not reproduce the suggested trend, we observe that
none of the parameter-free models presently available describes the trend 
in $G_M^n$ as observed in our data. The data show for the first time a 
significant violation of the scaling rule $G_M^p/\mu_p = G_M^n/\mu_n$.

\section{Conclusions and Outlook}
We have combined the ratio method proposed long ago 
with the possibilities offered only recently by the large duty
cycle electron beams to measure $G_M^n$ with substantially
decreased uncertainties.
The coincident detection of the electron and the knocked out nucleon 
reduces significantly uncertainties due to nuclear effects, 
whereas the large duty cycle allows to detect simultaneously 
protons and neutrons.

Recently, a new set of data has become available from the MAMI accelerator,
Mainz~\cite{thissymp} covering a similar range in momentum transfer. 
Although these data confirm the violation of the
above mentioned scaling rule, they differ significantly from our results.
This fact is under investigation, and a possible explanation has 
been given in section IIB. 

An experiment has been approved at the Bates accelerator, in which $G_M^n$
will be measured at small $Q^2$, i.e. $< 0.125$ (GeV/c)$^2$, using the
method as employed by Ref.~\cite{markowitz}. Although nuclear effects are
large for the proposed measurement, the result will be a continuous set
of data points up to the region where the existence of a sudden fluctuation 
in $G_M^n$ has been claimed~\cite{markowitz}.  

At TJNAF, an experiment has been approved in which $G_M^n$ will be measured 
using the ratio method at $Q^2 = 0.3 - 5.1$ (GeV/c)$^2$. One of the 
main experimental problems in this experiment will be the matching of the
proton and neutron phase spaces, due to the different techniques employed
to detect each of the nucleons.

\newpage

\begin{figure}
\strut\psfigure[4.0cm]{tof.psgood}
\caption{Time spectra of protons (line) and neutrons
(shaded area)  using a subset of the data from the measurement 
at $Q^2$=0.255 (GeV/c)$^2$.}
\label{figure:tof}
\end{figure}

\end{document}